# Inter-Domain Charge Transfer as a Rationale for Superior Photovoltaic Performances of Mixed Halide Lead Perovskites

Marine E. F. Bouduban,[a] Fabrizio Giordano,[b] Arnulf Rosspeintner,[c] Joël Teuscher,[a] Eric Vauthey,[c] Michael Grätzel,[b] and Jacques-E. Moser [a] *

[a] *Photochemical Dynamics Group, Institute of Chemical Sciences & Engineering, and Lausanne Centre for Ultrafast Science (LACUS), École polytechnique fédérale de Lausanne, CH-1015 Lausanne, Switzerland.*

[b] *Laboratory for Photonics and Interfaces, Institute of Chemical Sciences & Engineering, École polytechnique fédérale de Lausanne, CH-1015 Lausanne, Switzerland.*

[c] *Department of Physical Chemistry, University of Geneva, CH-1211 Geneva, Switzerland.*

\* *Corresponding author.* E-mail: je.moser@epfl.ch.

## Abstract

Organic-inorganic lead halide perovskites containing a mixture of iodide and bromide anions consistently perform better in donor-acceptor heterojunction solar cells than the standard methylammonium lead triiodide material. This observation is counterintuitive, as it is generally expected for photovoltaic materials that heterogeneities and compositional disorder cause reduced carrier diffusion length and conversion efficiency. Here, we combine ultrafast photoinduced electroabsorption and broadband fluorescence up-conversion spectroscopy measurements to scrutinize the carrier dynamics in mixed-cations, mixed-halide lead perovskite thin films. Our results evidence the formation of charge transfer excitons (CTE) across the boundaries of domains of various halide compositions. A global analysis of photoinduced transient Stark signals shows that CTE evolve gradually from Br-rich to I-rich domains over tens to hundreds of picoseconds. Rather than constituting recombination centres, boundaries between domains of various halide compositions appear then to favour charge carrier separation by driving photogenerated holes along channels of decreasing bromide content.

## Introduction

Since their inception as very promising photovoltaic materials, lead halide perovskites have attracted a vast scientific interest across the fields of chemistry, solid-state physics, photonics, and material sciences. Power conversion efficiencies achieved with simple, solution-processed donor-acceptor heterojunction devices based on this type of semiconductor now exceed 22%,[1,2] while very encouraging results were recently obtained in adressing the stability issue.[3,4]

The constituents of the perovskite materials have significantly evolved since the first devices made out of the standard methylammonium lead triiodide perovskite (MAPI, $CH_3NH_3PbI_3$) to the latest developments relying on mixed cations, mixed anions perovskite systems. Indeed, it clearly emerges, from the past two years, that mixed-composition perovskites consistently perform better than standard systems, via both a better short-circuit current, and a larger open-circuit voltage.[1,5-7] These parameters, commonly used to assess the performances of photovoltaic devices, directly relate to more fundamental properties of the material: the bandgap of the absorber, determined by its electronic structure, and the ratio between radiative and non-radiative charge carrier recombination quantum yields, which depends on more complex dynamical phenomena and scattering processes.

Non-radiative electron-hole recombination in semi-conductors usually occurs in presence of mid-bandgap states due to lattice imperfections or impurity atoms.[8] Inhomogeneous and disordered materials are also more likely to exhibit recombination centres, causing significant charge carrier losses and, concomitantly, decreasing their photovoltaic conversion efficiency.

The preparation of lead halide perovskites containing a mixture of organic methylammonium ($MA^+$) and formamidinium ($FA^+$), and inorganic cesium ($Cs^+$) and rubidium ($Rb^+$) cations, have shown to yield large monolithic grains, which dimensions can reach 1 μm, exceeding by far the typical 200 nm film thickness used in photovoltaic devices. The absence of grain boundaries along the transport path of the charge carriers improves significantly their diffusion length and must be beneficial to the photovoltaic conversion efficiency.[5,7,9,10]

Mixed halide materials containing both Br⁻ and I⁻ anions are quite attractive for LED application, due to the continuous tunability of their bandgap obtained by adjusting the Br⁻/I⁻ content ratio. These materials, however, appear problematic, as halide segregation tends to occur gradually in the dark [11] and under irradiation.[12-14] While this phase separation effect is believed to be attenuated in mixed-cation materials,[12] it remains to be understood which mechanism could make mixed halide perovskites exhibit enhanced photovoltaic conversion efficiencies compared to mixed cation lead triiodide, in spite of potential heterogeneities and an increased bandgap of the material that worsen the absorption mismatch with the solar spectrum.

Here, we used a combination of ultrafast transient absorption (TAS) and fluorescence up-conversion spectroscopy (FLUPS) techniques to study charge- and energy transfer processes taking place in transparent thin-films of lead halide perovskites of the composition $MA_yFA_{1-y}PbI_{3-x}Br_x$, containing a mixture of methylammonium ($MA^+$) and formamidinium ($FA^+$) cations, as well as mixed iodide ($I^-$) and bromide ($Br^-$) anions. The exploitation of the dynamics of transient photoinduced electroabsorption signals observed in TAS spectra allowed to evidence charge transfer between domains of various chemical compositions taking place in competition with energy transfer. These findings suggest that the separation of photogenerated carriers in regions characterized by a gradient of the valence band edge energy is at the origin of the improved efficiency of solar cells based on mixed halide perovskites.

## Results and discussion

Ultrafast transient absorption spectroscopy was applied to solution-processed mixed $MA_yFA_{1-y}PbI_{3-x}Br_x$ and standard $MAPbI_3$ perovskite thin films spin-coated on glass. The mixed-cations, mixed-halides perovskite material was prepared from solution reactants with an initial stoichiometry corresponding to the fractions $x = 0.45$ and $y = 0.15$. The mean composition of the final solid material is not expected to be significantly different. However, a substantial heterogeneity of bromide and iodide ions distribution is evidenced by the absorption spectrum of mixed perovskite films. Fig. 1 shows that mixed perovskite thin-films exhibit six resolved features at 454, 510, 560, 600 and 633 nm, which are believed to result from excitonic bands in domains characterized by different mixed halide compositions. On the other hand, standard $MAPbI_3$ perovskite samples exhibit the usual shape, with one feature at a probe wavelength $\lambda = 760$ nm, followed by a continuous increase in absorbance, on top of a second resolved feature at 480 nm.

The ultrafast transient absorption (TA) spectrum of the standard perovskite layer observed upon pulsed excitation of the sample at a pump wavelength $\lambda_{ex} = 390$ nm (Fig. 2A) exhibits a large negative feature at a probe wavelength $\lambda = 760$ nm, assigned to a combination of ground-state bleaching and stimulated emission, a broad positive band at shorter wavelengths and, finally, a second negative band at 480 nm, which assignment remains under discussion.[15-21] The transient spectrum of the mixed perovskite recorded in identical conditions reveals a similar negative feature centred at 750 nm (Fig. 2B). The latter, however, appears strikingly different, with strong oscillations in the $\lambda = 500$-630 nm region. The latter peculiar features are only observed in the mixed perovskite samples, where they are not modified by the film thickness, nor by the light angle of incidence. The occurrence of spectral artefacts due to interferences can, therefore, be excluded. In the following, we focus on explaining the complexity of the obtained spectra that hints at the presence of more dynamical processes or photo-excited populations than in the standard material case.

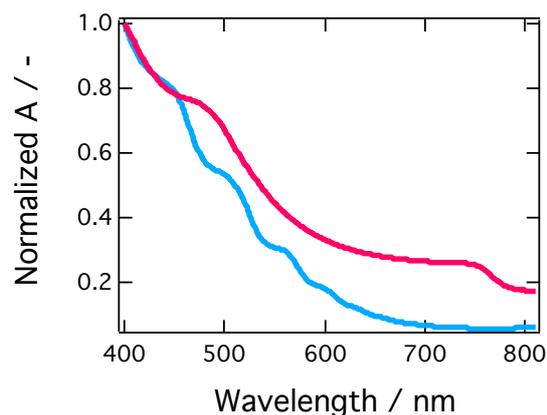

**Figure 1.** Absorbance spectra of standard $MAPbI_3$ (magenta) and mixed-composition $MA_yFA_{1-y}Br_xI_{3-x}$ (blue) perovskite film samples measured in transmittance mode.

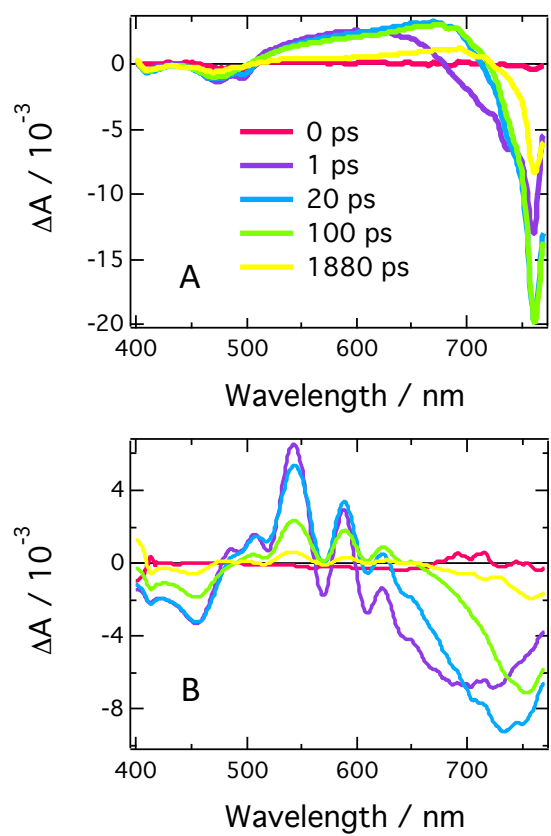

**Figure 2.** Ultrafast TA spectra recorded upon pulsed photoexcitation of perovskite thin films at a pump wavelength $\lambda_{ex} = 390$ nm and various probe time delays: red: 0 ps, purple: 1 ps, blue: 5 ps, green: 100 ps and yellow: 1880 ps. A) $MAPbI_3$, pump energy fluence 7 μJ cm⁻². B) $MA_yFA_{1-y}PbI_{3-x}Br_x$, pump fluence 35 μJ cm⁻².

As a first step, we consider the oscillations observed in the spectral region between 500 and 630 nm in Fig. 2B, which we assign to a photoinduced electroabsorption signal.[22,23] This feature arises from the photogeneration of electron-hole pairs that induce local electric fields, affecting the surrounding material and yielding changes in its absorption properties. Depending on the type of sample under study, and more specifically, on the level of correlation of the charge carriers, the electroabsorption can be modelled in different ways: Stark effect, Franz-Keldysh effect, or excitonic electroabsorption theories.[24]

Methylammonium lead triiodide perovskite exhibits low exciton binding energy and, hence, low correlation between charge carriers.[24-31] As such, its electroabsorption signal has been shown to correspond to the low-field Franz-Keldysh-Aspnes effect.[32] On the other hand, the exciton binding energy of hybrid lead halide perovskite increases as one moves up along the halogen column of the periodic table.[24,27,30,33] As a consequence, we assume that the electroabsorption of perovskite layers involving mixed $Br^-/I^-$ anions can be modelled reasonably well within the Stark theory. In such a framework, the electroabsorption signal can be shown to correspond to a linear combination of the first and second derivatives of the linear absorbance spectrum (see ESI†). For isotropic samples, the related amplitudes of these two components are given respectively by the change in the polarizability of the material and by the change in the dipole moment upon the electronic transition of interest.

Figure 3 displays the transient absorbance spectrum of a thin film of $MA_yFA_{1-y}PbI_{3-x}Br_x$ mixed perovskite measured at a probe time delay of 0.2 ps, together with the first and second derivatives of the linear absorbance spectrum of the same sample (Fig. 1). Both derivatives exhibit broadband oscillations, and the TA signal appears to be clearly dominated by a second-derivative contribution. This implies the occurrence of a dipole moment change upon photoexcitation. Such a dipole moment change can only arise from charge-transfer excitons (CTE), namely electrostatically-bound electron-hole pairs, where both photogenerated carriers are separated by a junction or a domain boundary.[34] As a consequence, the second derivative-dominated line-shape of the observed photoinduced electroabsorption signal, in the framework of Stark theory, constitutes an evidence of the generation of CTE at boundaries between grains of homogeneous composition, or at junctions between hetero-domains. From Figs 1 and 3, it appears clearly that the latter is the most likely conclusion, as the CTE signature spans a broad spectral region, from 500 nm to 630 nm, where several shoulders are observed in the absorption spectrum that are likely to correspond to excitonic bands of domains characterized by various degrees of halide mixing. We dismiss here the possibility of light-induced phase segregation to stand at the origin of this observation, as no sample modification, nor aging occurred during the measurements. We therefore conclude that the mixed halide perovskite film is intrinsically constituted of heterogeneous domains, at the boundaries of which photogenerated electron-hole pairs get trapped in the form of charge transfer excitons.

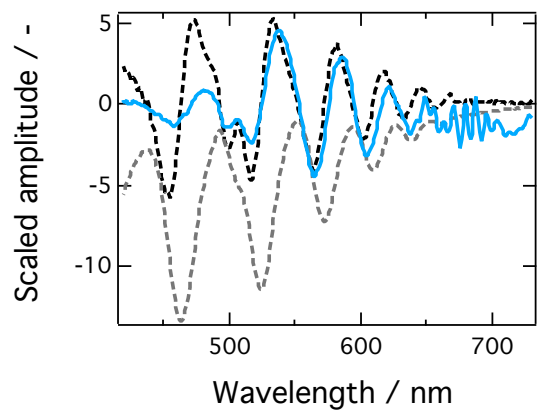

**Figure 3.** Superposition of the TA spectrum of the mixed perovskite sample measured at a probe time delay of 0.2 ps (blue line), with the first and second derivatives of the linear absorbance spectrum (grey and black dotted lines, respectively). The transient spectrum was obtained upon photoexcitation at $\lambda_{ex}$ = 390 nm and at a pump energy fluence of 20 µJ cm$^{-2}$.

The valence band maximum of lead halide perovskites is determined by the nature of the halide, due to the change from 4p (for $Br^-$) to 5p (for $I^-$) valence orbitals. As a result, the bandgap of the material is directly affected by the $Br^-/I^-$ content ratio, while the effect of the cation proportion is limited. We can, therefore, go one step further by explaining the compositional range suggested by the oscillations of the electroabsorption signal: The first clear oscillation peaks at 530 nm correspond to the absorption maximum of $MA_yFA_{1-y}PbBr_3$ for a large span of $y$ fraction values. The last observable peak on Figure 2B is centred at 670 nm, previously assigned to a $Br_{1.5}I_{1.5}$ halide mixture.[35] We note also that the amplitude of the oscillations decreases as one moves towards smaller $Br^-/I^-$ ratio, in the lower energy part of the spectrum. This is expected to be due to the decreased exciton binding energy of iodide rich materials, translating into a smaller dipole moment change and a weaker CTE-dominated Stark signal.

The presence of heterogeneous domains within mixed halide perovskite films raises questions about the way charge carriers travel through the material. In other words, it questions the nature of the interaction between those domains and its impact on charge separation and transport, a key process for device efficiencies, potentially accounting for the better performance of mixed perovskite-based solar cells. In such a context, we wish to decompose our time-series of transient absorbance spectra in a linear combination of the important processes underlying it. Those processes can then be identified and one can assess their role in the fate of charge carriers within the material. A straightforward way to

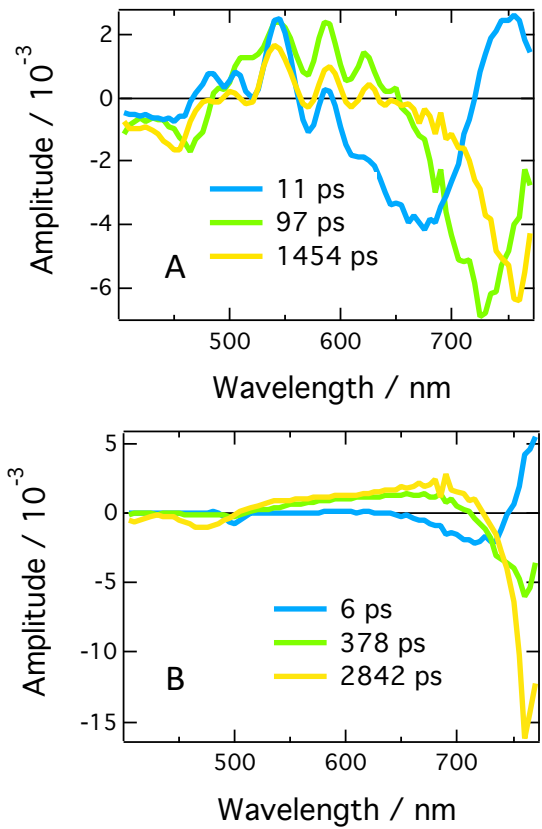

**Figure 4.** Results of the global fitting procedure (tri-exponential): Decay associated spectra (DAS). A) MA$_y$FA$_{1-y}$PbI$_{3-x}$Br$_x$. Blue: 11 ps, green: 97 ps and yellow: 1454 ps. B) MAPbI$_3$. Blue: 6 ps, green: 378 ps and yellow: 2842 ps.

___

achieve this is to resort to a global analysis of our time-resolved spectral data, which consists in considering our system as constituted of independent species, each evolving exponentially in time. The outcomes of such a global analysis procedure are the decay associated spectra (DAS) of each of the identified processes. By design, each DAS is associated with a time-constant that represents the decay time of the process underlying the spectral signature under consideration.[36]

The results of the global fitting procedure of the data presented in Fig. 2 are displayed in Fig. 4. Details of the multi-exponential fitting procedure are provided in the ESI† along with sample results. In addition, Fig. S1 (ESI†) assesses the fitting quality by displaying it on top of experimental points. For consistence and comparison purposes, both sets of data have been fitted with three exponentials starting 2 ps after pulsed photoexcitation.

When looking at Figure 4A, it is clear that the resulting three DAS correspond well to the data from Figure 2B, revealing a good fit quality. It also emerges that all three DAS exhibit the oscillatory contribution discussed above, assigned to the presence of CTE. In addition, the DAS feature a large negative component that describes the photoexcited population itself (decay of the bleaching signal). This feature shifts from 670 nm at 11 ps to 720 nm at 97 ps, to settle at 750 nm at longer time-delays, i.e. 1500 ps. The three DAS can thus be assigned as follows: The 6 ps component involves contributions from the short-time phenomena (bandgap renormalisation and carrier thermalisation), together with the population decay arising from CTE-mediated transfer from a large bandgap (Br-rich) domain to an intermediate bandgap domain. The second DAS (378 ps) represents CTE-mediated transfer from the latter intermediate bandgap domain to the lowest bandgap one. In turn, the last component features the decay of the equilibrium population from the latter composition. We can here be even more specific by assigning those various bandgaps to a given Br$^-$/I$^-$ ratio, within the assumption of a fixed MA, FA proportion or a negligible effect of the cation: In this context, the species at 670 nm, 720 nm and 750 nm correspond respectively to the halide compositions $x = 1.5$, $x = 1$, and $x = 0.5$.[35] Other intermediate compositions, such as the ones highlighted in Fig. 2, might as well exist, but are not resolved in this study.

Moving to Figure 4B, featuring the global analysis result for the standard MAPbI$_3$ perovskite, we see that the three obtained DAS differ significantly from the mixed halide case. Again, the first DAS involves the aforementioned short-time phenomena, while the second and third DAS represent the decay of the photoexcited

___

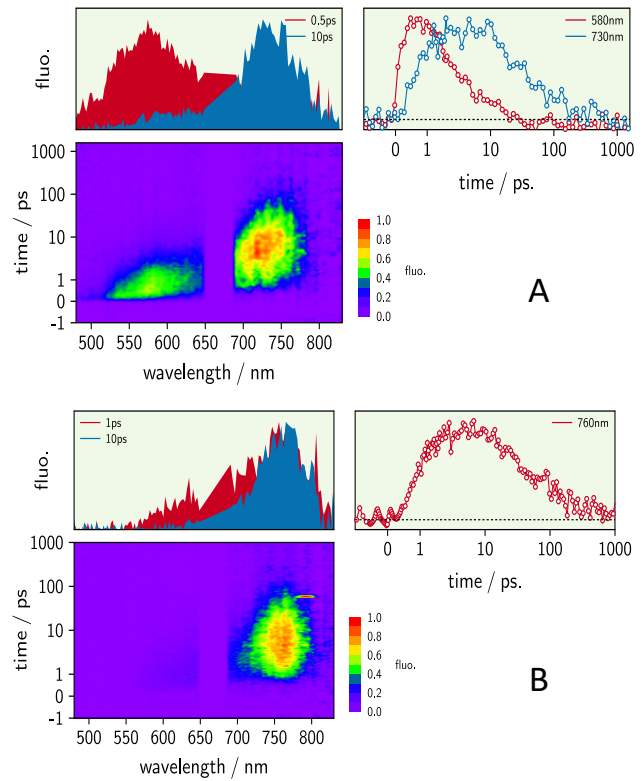

**Figure 5.** Fluorescence up-conversion measurements of insulated thin-films of MA$_y$FA$_{1-y}$PbI$_{3-x}$Br$_x$ upon pulsed photoexcitation at a pump wavelength $\lambda_{ex} = 400$ nm. A) Pulse excitation energy: 200 nJ. B) Pulse excitation energy: 50 nJ. For both A and B, bottom left panels feature the 3D data, with (normalized) luminescence intensities indicated as a color scale. Top left panels represent spectra at selected times, and top right panels display the dynamics extracted at the maxima of the emission peaks.

population via different mechanisms, such as radiative and non-radiative recombination, that cannot be distinguished with our technique.[40] No sign of population transfer is visible, in this case, which is compatible with a homogeneous material.

To support the conclusion of inter-domain charge transfer in mixed perovskite materials, fluorescence up-conversion measurements at two different excitation energies have been performed on MA$_y$FA$_{1-y}$PbI$_{3-x}$Br$_x$ perovskite films and are displayed in Figure 5. Fig. 5A, featuring the data obtained at the highest excitation energy (200 nJ), shows that the mixed halide perovskite layer exhibits multisite emission, with two peaks centred at 570 and 780 nm, respectively, and generally evolving towards the red. The dynamics of those peaks differ but are linked: The 570 nm signal rises quickly (within the instrument response function), and its decay takes place on a timescale that corresponds to the rise of the 780 nm peak. Coexistence of different perovskite compositions, which are interacting within our mixed perovskite layer is evidenced by this shift in emission energy over time.

Fig. 5B shows the same sample at a lower excitation energy and it emerges that most of the emitted light is here centred at 770 nm, with a greatly reduced contribution from the blue part of the spectrum at early times. This observation is in turn compatible with a kinetic competition of the 570 nm emission with another deactivation channel, favoured at lower carrier densities. This particular process is assigned to charge transfer towards smaller bandgap neighbouring domains, in agreement both with the related dynamics in Fig. 5A and with the global fitting results.

In addition, knowing that the nature of the halide mostly affects the valence band, we state that the carrier under observation in those charge transfers is the hole, and not the electron.[12,41,42]

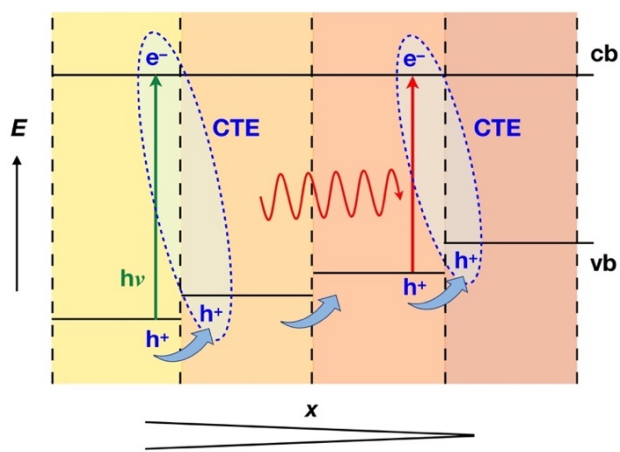

**Figure 6.** Cartoon illustrating domains in the MA$_y$FA$_{1-y}$PbI$_{3-x}$Br$_x$ perovskite, whose bromide fraction $x$ is decreasing from left to right. Charge transfer excitons (CTE) formed at domain boundaries (dashed ellipses) mediate the transport of photogenerated holes in the same direction. CTE formation competes kinetically with energy transfer to narrower bandgap domains (wavy red arrow).

The occurrence of selective hole transfer in mixed perovskite systems, therefore, is likely to play a key role in the enhanced performance of MA$_y$FA$_{1-y}$PbI$_{3-x}$Br$_x$ and other mixed composition perovskite solar cells.[1,5,7] Indeed, the ability to selectively transfer a charge carrier across a boundary ensures an efficient charge separation within the perovskite layer, and less losses due to trap-assisted recombination that are vertical in the k-space. We propose that the subsequent transport of the separated hole occurs through percolation across domains with progressively lower Br⁻ content.[43,44] This sets the focus on the importance of the optimization of the relative proportion of iodide and bromide domains, as the percolation threshold must be reached for the extraction of charge carriers towards their selective extracting layers to occur. The model proposed here is illustrated by Figure 5. Similar transfer processes and their beneficial effects have already been reported for different types of perovskite systems, either as a cascade transfer between nanoparticles or in the form of photon recycling in thin films.[43-45]

**Table 1.** JV-curve electrochemical characterization of perovskite solar cells. The films were prepared in the same way as for the spectroscopic measurements. $J_{SC}$ is the short-circuit current density, $V_{OC}$ the open-circuit voltage, $FF$ the fill-factor, $PCE$ the incident solar-to-electrical power conversion efficiency, and $I_{light}$ the incident light intensity.

| Perovskite | Scan direction | $J_{sc}$ mA cm$^{-2}$ | $V_{OC}$ V | $FF$ | $PCE$ % | $I_{light}$ mW cm$^{-2}$ |
|---|---|---|---|---|---|---|
| Standard | backward | -18.24 | 1.07 | 0,72 | 14.4 | 97.9 |
|  | forward | -18.24 | 1.07 | 0,62 | 12.4 |  |
| Mixed | backward | -18.13 | 1.09 | 0,76 | 15.3 | 97.9 |
|  | forward | -18.13 | 1.08 | 0,73 | 14.6 |  |

To show that TAS and FLUPS measurements and the conclusions drawn from them are indeed relevant to real functional solar cells, mixed perovskite films prepared exactly in the same conditions were incorporated in devices and their photovoltaic performances compared to those of the standard MAPbI$_3$. JV-curve measurements are presented in Figs. S3 and S4 (ESI†) and the results are summarized in Table 1. As expected, mixed perovskite films exhibit a better power conversion efficiency, as well as reduced hysteresis. Regarding the absolute PCE figures, one has to remember that the measured films are quite thin and do not allow in particular complete harvesting of the incident light.

## Conclusions

Ultrafast transient absorption spectra of MA$_y$FA$_{1-y}$PbI$_{3-x}$Br$_x$ thin-films appear quite different from those obtained from the standard MAPbI$_3$ material. Large oscillations are observed for the mixed halide perovskite samples in the visible range. These oscillations are characteristic of a photoinduced transient Stark effect and are fitted by the second derivative of the ground state absorbance spectrum of the material, which features several excitonic bands of domains of various mixed halide composition. The transient Stark signal asserts the formation of permanent dipoles upon photoexcitation, corresponding to charge transfer excitons (CTE) generated astride the boundaries of domains in the bulk of the semiconductor.

A global analysis of transient absorbance data confirms that the mixed halide perovskite films contain heterogeneous domains characterized by different degrees of halide mixing. These exhibit different bandgap energies and excitonic peak positions, assigned to different mixed perovskite stoichiometries. The photogenerated carrier population in each of those domains furthermore undergoes CTE-mediated hole transfer from the largest (Br-rich) to the lowest (I-rich) bandgap domains. This scenario is supported by fluorescence up-conversion measurements of $MA_yFA_{1-y}PbI_{3-x}Br_x$ samples: Multisite emission is observed, assessing the presence of various radiative recombination energies, which confirms the occurrence of hetero-domains with different energetics. Alternatively, at lower excitation density, the main portion of the radiative relaxation occurs at 770 nm, corresponding to the steady-state emission wavelength. This demonstrates the presence of competing deactivation pathways, namely radiative recombination, statistically favoured at high fluence, and charge transfer towards neighbouring domains, dominating at low fluence.

We furthermore suggest that, following their transfer, holes percolate along channels of decreasing bromide content until they reach their selective extracting layer. A random distribution of hetero-domains within the film implies that only a minority of photogenerated carriers are actually concerned by this effect. Still, such a mechanism for selective hole transfer across the active layer enhances the efficiency of charge separation and accounts for the superior performances of mixed halide perovskite systems.

Our observations open the door to specific active layer engineering to achieve an optimum charge separation process. They suggest in particular that a gradient of the halide composition could be used to selectively drive the holes towards the iodide-rich side placed at the contact of the hole transporting material.

## Experimental section

### Samples preparation

Two different sample architectures were used, which were adapted to the spectroscopic techniques employed: Perovskite/glass and insulating layer/perovskite/insulating layer/gold samples were studied with ultrafast transient absorption spectroscopy (TAS) in transmission mode and broadband fluorescence up-conversion spectroscopy (FLUPS), respectively.

The samples were prepared according to previously reported protocols with the exception of the concentration of the perovskite precursor solution which was kept at 0.7 M in order to obtain a thinner layer of the absorber material to facilitate the optical and spectroscopy measurements.[43,44] The same precursor concentration was used to prepare complete solar cell devices, to ensure a direct correlation between the spectroscopic characterization and the photovoltaic results.

Nippon Sheet Glass 10 Ω/sq substrates (for the fluorescence up-conversion spectroscopy measurements), respectively microscope glass (transient absorbance spectroscopy experiments) were cleaned by ultra-sonication in a bath of 2% Hellmanex aqueous solution for 30 minutes. After rinsing with deionised water and ethanol, the substrates were further cleaned with UV ozone treatment for 15 min.

The perovskite films were deposited on the substrate from a precursor solution containing $MA_{0.15}FA_{0.85}PbI_{2.55}Br_{0.45}$, on the one hand, and standard $MAPbI_3$, on the other hand, in anhydrous DMF:DMSO 4:1 (v:v) at a concentration of 0.7 M. The perovskite solution was spin-coated in a two steps program at 1000 and 6000 rpm for 10 and 30 s, respectively. During the second step, 120 μL of chlorobenzene were poured on the spinning substrate 10 s prior to the end of the program. The substrates were then annealed at 100°C for 1 hour in a dry air-filled glove box.

While the initial stoichiometry of the reactants used to prepare the mixed perovskite samples was precisely defined, the final composition of the material after crystallization is believed to be rather heterogeneous. Hence, the mixed cation, mixed halide perovskite material under study will be designated here by the formula $MA_yFA_{1-y}PbI_{3-x}Br_x$.

The insulating layers for the FLUPS samples were constituted by a first coating of pure poly(methylmethacrylate) (PMMA) and a second layer of PMMA and alumina nanoparticles. Such insulating double layer was applied on bare FTO and on top of the perovskite. The PMMA layer was deposited by spin-coating for 20 s at 4000 rpm with a ramp of 2000 rpm, using a solution of 20 mg/mL of PMMA solution in toluene. The PMMA/ alumina mixed layer was deposited by spin-coating for 20 s at 4000 rpm with a ramp of 2000 rpm. The dispersion of alumina nanoparticles (Aldrich) was diluted 10:1 in volume with a PMMA solution in toluene (3 mg/ mL). Finally, to make sure the samples intended to be studied with FLUPS were properly working devices, a small gold electrode (120 mm-thick) was thermally evaporated under high vacuum onto each of them, allowing for photovoltaic characterization. The subsequent FLUPS measurements were carried out on the electrode-less area of the samples, in transmittance mode.

### Photovoltaic devices fabrication and testing

Complete solar cell devices were prepared by depositing the perovskite solution precursor onto a FTO substrate (Nippon Sheet Glass 10 Ω/sq) comprising a 15 nm thin coating of $SnO_2$ deposited by ALD as electron transporting layer. After annealing the perovskite layer, 100 nm of *spiro*-MeOTAD was spun to form the hole transporting material (HTM) layer. The HTM was doped at a molar ratio of 0.5, 0.03 and 3.3 with bis(trifluoromethylsulfonyl)-imide lithium salt (Li-TFSI, Sigma Aldrich), tris(2-(1H-pyrazol-1-yl)-4-tert-butylpyridine)-cobalt(III) (FK209, Dyenamo), and 4-tert-butylpyridine (TBP, Sigma Aldrich), respectively. Finally, a 70-80 nm-thick gold electrode was thermally evaporated on top of the HTM layer. Test solar cell elements had a 5 × 5 mm square geometry.

Solar cells merit parameters were measured using a 300 W xenon light source (Ushio). The spectral mismatch between AM1.5G and the simulated illumination was reduced by the use of a AM1.5G filter (Newport). The light intensity was calibrated with a Si photodiode equipped with an IR-cutoff filter (KG3, Schott) and it was recorded prior to each measurement. Current-voltage characteristics of the cells were obtained by applying an external voltage bias, while measuring the current response with a digital source meter (Keithley 2400). The voltage scan rate was 10 mV s$^{-1}$. The cell area was defined by a 4 × 4 mm (0.16 cm$^2$) black metal mask.

**Spectroscopic methods**

Ultrafast transient absorbance (TA) spectra of standard and mixed perovskites thin-films were acquired using femtosecond pump-probe spectroscopy with a pump wavelength $\lambda_{ex}$ = 390 nm. The pump beam was obtained by frequency doubling the output of a chirped pulse amplified Ti:Sapphire laser (CPA-2001, Clark-MXR, 778 nm fundamental central wavelength, 120 fs pulse duration, 1 kHz repetition rate) in a BBO crystal, yielding 200 fs pulses. The probe beam was generated by directing a portion of the 778 nm fundamental output of the laser into a CaF$_2$ crystal, yielding a white light continuum measured over a 400-780 nm spectral domain. The probe fluence at the sample was much lower than that of the pump (7-35 μJ cm$^{-2}$). Similarly, the diameter of the probe beam was smaller to ensure homogeneous excitation of the probed area. The dynamics of the photo-induced signals were obtained with a computer-controlled delay-line on the pump path. The probe beam was split before the sample into a beam going through the sample (signal beam) and a reference beam. Both signal and reference beams were directed to respective spectrographs (Princeton Instruments, Spectra Pro 2150i) and detected pulse-to-pulse with 512×58 pixels back-thinned CCD cameras (Hamamatsu S07030-0906). The pump beam was chopped at half of the laser frequency (500 Hz) and a satisfying signal-to-noise ratio was obtained by typically averaging 3,000 spectra. The time resolution of the experiment was 250 fs.[41]

Femtosecond time-resolved broadband fluorescence up-conversion spectra (FLUPS) were measured with a previously described instrument.[45] The setup had a time-resolution of 170 fs (FWHM of the instrument response function). The experimental spectra were corrected for the wavelength-dependent detection sensitivity, using a set of secondary emissive standards (covering the spectral range 415-850 nm), and for the temporal chirp, using the wavelength-independent instantaneous response of 2,5-bis(5-tert-butyl-benzoxazol-2-yl)thiophene (BBOT) in acetonitrile. The samples were not moved during the measurement. Integration times per time-delay were in the range 2-4 s, with each time-scan being averaged 10 times.

**Acknowledgements**

Financial support by the Swiss National Science Foundation (SNSF, grant no. 20001_175729) and the National Center of Competence in Research "Molecular Ultrafast Science and Technology" (NCCR-MUST), a research instrument of the SNSF, is gratefully acknowledged.

**References**

1. D. Bi, W. Tress, M. I. Dar, P. Gao, J. Luo, C. Renevier, K. Schenk, A. Abate, F. Giordano, J. P. Correa Baena, J. D. Decoppet, S. M. Zakeeruddin, M. K. Nazeeruddin, M. Grätzel and A. Hagfeldt, *Sci. Adv.*, 2016, **2**, e1501170.
2. W. S. Yang, B.-W. Park, E. H. Jung, N. J. Jeon, Y. C. Kim, D. U. Lee, S. S. Shin, J. Seo, E. K. Kim, J. H. Noh and S. I. Seok, *Science*, 2017, **356**, 1376.
3. N. Arora, M. I. Dar, A. Hinderhofer, N. Pellet, F. Schreiber, S. M. Zakeeruddin and M. Grätzel, *Science*, 2017, **358**, 768.
4. G. Grancini, C. Roldán-Carmona, I. Zimmermann, E. Mosconi, X. Lee, D. Martineau, S. Narbey, F. Oswald, F. De Angelis, M. Grätzel and M. K. Nazeeruddin, *Nat. Commun.*, 2017, **8**, 15684.
5. M. Saliba, T. Matsui, J.-Y. Seo, K. Domanski, J.-P. Correa-Baena, M. K. Nazeeruddin, S. M. Zakeeruddin, W. Tress, A. Abate, A. Hagfeldt and M. Grätzel, *Energy Environ. Sci.*, 2016, **9**, 1989.
6. W. Tress in *Organic- Inorganic Halide Perovskite Photovoltaics*, Eds N.-G. Park, M. Grätzel and T. Miyasaka, Springer, New York, 2016.
7. M. Saliba, T. Matsui, K. Domanski, J.-Y. Seo, A. Ummadisingu, S. M. Zakeeruddin, J. P. C. Baena, W. Tress, A. Abate, A. Hagfeldt and M. Grätzel, *Science*, 2016, **354**, 206.
8. V. N. Abakumov, V. I. Perel and I. N. Yassievich, *Nonradiative Recombination in Semiconductors*, Volume 33, Elsevier, Amsterdam, 1991.
9. A. A. Paraecattil, J. De Jonghe-Risse, V. Pranculis, J. Teuscher and J. E. Moser, *J. Phys. Chem. C*, 2016, **120**, 19595.
10. J. Teuscher, J. C. Brauer, A. Stepanov, A. Solano, A. Boziki, M. Chergui, J.-P. Wolf, U. Rothlisberger, N. Banerji and J. E. Moser, *Struct. Dyn.*, 2017, **4**, 061503.
11. A. Sadhanala, F. Deschler, T. H. Thomas, S. E. Dutton, K. C. Goedel, F. C. Hanusch, M. L. Lai, U. Steiner, T. Bein, P. Docampo, D. Cahen and R. H. Friend, *J. Phys. Chem. Lett.*, 2014, **5**, 2501.
12. S. J. Yoon, S. Draguta, J. S. Manser, O. Sharia, W. F. Schneider, M. Kuno and P. V. Kamat, *ACS Energy Lett.*, 2016, **1**, 290.
13. D. J. Slotcavage, H. I. Karunadasa and M. D. McGehee, *ACS Energy Lett.*, 2016, **1**, 1199.
14. A. J. Barker, A. Sadhanala, F. Deschler, M. Gandini, S. P. Senanayak, P. M. Pearce, E. Mosconi, A. J. Pearson, Y. Wu, A. R. Srimath Kandada, T. Leijtens, F. De Angelis, S. E. Dutton, A. Petrozza and R. H. Friend, *ACS Energy Lett.*, 2017, **2**, 1416.
15. K. G. Stamplecoskie, J. S. Manser and P. V. Kamat, *Energy Environ. Sci.*, 2014, **8**, 208.
16. D. W. deQuilettes, S. M. Vorpahl, S. D. Stranks, H. Nagaoka, G. E. Eperon, M. E. Ziffer, H. J. Snaith and D. S. Ginger, *Science*, 2015, **348**, 683.
17. J. S. Manser, B. Reid and P. V. Kamat, *J. Phys. Chem. C*, 2015, **119**, 17065.
18. G. Xing, N. Mathews, S. Sun, S. S. Lim, Y. M. Lam, M. Grätzel, S. Mhaisalkar and T. C. Sum, *Science*, 2013, **342**, 344.
19. P. Piatkowski, B. Cohen, F. Javier Ramos, M. Di Nunzio, M. K. Nazeeruddin, M. Grätzel, S. Ahmad and A. Douhal, *Phys. Chem. Chem. Phys.*, 2015, **17**, 14674.

20 M. B. Price, J. Butkus, T. C. Jellicoe, A. Sadhanala, A. Briane, J. E. Halpert, K. Broch, J. M. Hodgkiss, R. H. Friend and F. Deschler, *Nat. Commun.*, 2015, **6**, 1–8.
21 T. C. Sum and N. Matthews, *Energy Environ. Sci.*, 2014, **7**, 2518.
22 M. T. Trinh, X. Wu, D. Niesner and X. Y. Zhu, *J. Mater. Chem. A*, 2015, **3**, 9285.
23 V. Roiati, E. Mosconi, A. Listorti, S. Colella, G. Gigli and F. De Angelis, *Nano Lett.*, 2014, **14**, 2168.
24 I. B. Koutselas, L. Ducasse and G. C. Papavassiliou, *J. Phys.: Condens. Matter*, 1996, **8**, 1217.
25 K. Tanaka, T. Takahashi, T. Ban, T. Kondo, K. Uchida and N. Miura, *Solid State Commun.*, 2003, **127**, 619.
26 T. J. Savenije, C. S. Ponseca Jr., L. Kunneman, M. Abdellah, K. Zheng, Y. Tian, Q. Zhu, S. E. Canton, I. G. Scheblykin, T. Pullerits, A. Yartsev and V. Sundström, *J. Phys. Chem. Lett.*, 2014, **5**, 2189.
27 M. Saba, F. Quochi, A. Mura and G. Bongiovanni, *Acc. Chem. Res.*, 2015, 166.
28 A. Miyata, A. Mitioglu, P. Plochocka, O. Portugall, J. T.-W. Wang, S. D. Stranks, H. J. Snaith and R. J. Nicholas, *Nat. Phys.*, 2015, **11**, 582.
29 A. M. Soufiani, F. Huang, P. Reece, R. Sheng, A. Ho-Baillie and M. A. Green, *Appl. Phys. Lett.*, 2015, **107**, 231902.
30 Y. Yang, M. Yang, Z. Li, R. Crisp, K. Zhu and M. C. Beard, *J. Phys. Chem. Lett.*, 2015, **6**, 4688.
31 K. Galkowski, A. Mitioglu, A. Miyata, P. Plochocka, O. Portugall, G. E. Eperon, J. T.-W. Wang, T. Stergiopoulos, S. D. Stranks, H. J. Snaith and R. J. Nicholas, *Energy Environ. Sci.*, 2016, **9**, 962.
32 M. E. Ziffer, J. C. Mohammed and D. S. Ginger, *ACS Photonics*, 2016, **3**, 1060.
33 N. Sestu, M. Cadelano, V. Sarritzu, F. Chen, D. Marongiu, R. Piras, M. Mainas, F. Quochi, M. Saba, A. Mura and G. Bongiovanni, *J. Phys. Chem. Lett.*, 2015, **6**, 4566.
34 G. Lanzani, *The Photophysics behind Photovoltaics and Photonics*, Wiley-VCH, Weinheim, 2012.
35 T. J. Jacobsson, J.-P. Correa-Baena, M. Pazoki, M. Saliba, K. Schenk, M. Grätzel and A. Hagfeldt, *Energy Environ. Sci.*, 2016, **9**, 1706.
36 I. H. M. van Stokkum, D. S. Larsen and R. van Grondelle, *Biochim. Biophys. Acta, Bioenerg.*, 2004, **1657**, 82.
37 G. Grancini, A. R. S. Kandada, J. M. Frost, A. J. Barker, M. De Bastiani, M. Gandini, S. Marras, G. Lanzani, A. Walsh and A. Petrozza, *Nat. Photon.*, 2015, **9**, 695.
38 F. Brivio, K. T. Butler, A. Walsh and M. van Schilfgaarde, *Phys. Rev. B*, 2014, **89**, 155204.
39 J. Even, L. Pedesseau, C. Katan, M. Kepenekian, J.-S. Lauret, D. Sapori and E. Deleporte, *J. Phys. Chem. C,* 2015, **119**, 10161.
40 P. Gratia, G. Grancini, J.-N. Audinot, X. Jeanbourquin, E. Mosconi, I. Zimmermann, D. Dowsett, Y. Lee, M. Grätzel, F. De Angelis, K. Sivula, T. Wirtz and M. K. Nazeeruddin, *J. Am. Chem. Soc.*, 2016, **138**, 15821.
41 M. E. F. Bouduban, A. Burgos-Caminal, R. Ossola, J. Teuscher and J. E. Moser, *Chem. Sci.*, 2017, **8**, 4371.
42 L. M. Pazos-Outòn, M. Szumilo, R. Lamboll, J. M. Richter, M. Crespo-Quesada, M. Abdi-Jalebi, H. J. Beeson, M. Vrućinić, M. Alsari, H. J. Snaith, B. Ehrier, R. H. Friend and F. Deschler, *Science*, 2016, **351**, 1430.
43 J. P. C. Baena, L. Steier, W. Tress, M. Saliba, S. Neutzner, T. Matsui, F. Giordano, T. J. Jacobsson, A. R. S. Kandada, S. M. Zakeeruddin, A. Petrozza, A. Abate, M. K. Nazeeruddin, M. Grätzel and A. Hagfeldt, *Energy Environ. Sci.*, 2015, **8**, 2928.
44 F. Giordano, A. Abate, J. P. C. Baena, M. Saliba, T. Matsui, S. H. Im, S. M. Zakeeruddin, M. K. Nazeeruddin, A. Hagfeldt and M. Graetzel, *Nat. Commun.*, 2016, **7**, 10379.
45 17 M. Gerecke, G. Bierhance, M. Gutmann, N. P. Ernsting and A. Rosspeintner, *Rev. Sci. Instr.*, 2016, **87**, 053115.

# Inter-Domain Charge Transfer as a Rationale for Superior Photovoltaic Performances of Mixed Halide Lead Perovskites

Marine E. F. Bouduban,[a] Fabrizio Giordano,[b] Arnulf Rosspeintner,[c] Joël Teuscher,[a] Eric Vauthey,[c] Michael Grätzel,[b] and Jacques-E. Moser [a] *

[a.] *Photochemical Dynamics Group, Institute of Chemical Sciences & Engineering, and Lausanne Centre for Ultrafast Science (LACUS), École polytechnique fédérale de Lausanne, CH-1015 Lausanne, Switzerland.*

[b.] *Laboratory for Photonics and Interfaces, Institute of Chemical Sciences & Engineering, École polytechnique fédérale de Lausanne, CH-1015 Lausanne, Switzerland.*

[c.] *Department of Physical Chemistry, University of Geneva, CH-1211 Geneva, Switzerland.*

* Corresponding author. E-mail: je.moser@epfl.ch.

# Supplementary Information

## 1. Electroabsorption

From the perturbation theory, one can define the electroabsorption (EA) signal as a perturbative expansion in the electric field :[1]

$$\Delta A = -\frac{\partial A}{\partial \lambda} \cdot \boldsymbol{m}_{0k}\boldsymbol{E} - \frac{1}{2}\frac{\partial A}{\partial \lambda} \cdot \boldsymbol{p}_{0k}\boldsymbol{E}^2 + \frac{1}{2}\frac{\partial^2 A}{\partial \lambda^2} \cdot (\boldsymbol{m}_{0k}\boldsymbol{E})^2 + ... \quad (S1)$$

where $A$ is the absorbance and $\boldsymbol{E}$ the electric field vector. $\boldsymbol{m}_{0k}$ and $\boldsymbol{p}_{0k}$ correspond respectively to the permanent dipole moment and polarizability changes upon the transition of interest (0 → k). Note that the first term (linear in the field) is expected to cancel out for isotropic samples.

As it clearly emerges from Eq. S1, the EA signal can be decomposed into a linear combination of first and second derivatives of the linear absorption spectrum, with the amplitude of the latter contribution provided by the change in the permanent dipole moment.

## 2. Global Analysis

Global fits of transient absorbance data have been used in this work to assess the presence of inter-domain interactions. A comprehensive review of global analysis can be found elsewhere,[2,3] and an illustration in the case of transient absorption spectroscopy can be found in references [4].

In the present case, the choice between a parallel or sequential model was tricky. We selected the former, as some parts of the TA signal actually arise from a linear combination of individual processes (such as the bleaching dynamics, or the dynamics of the broad positive signal between 600-700 nm). However, we did not go as far as applying an actual kinetic model but focused on extracting the spectral range of individual dynamical processes.

To perform a global fit, data must first be sampled, and an appropriate number of kinetic traces must be extracted. Then, the point is to apply a fitting equation, usually multiexponential, where all the traces are forced to evolve with the same time constant(s). This yields the following point by point procedure:

1) Extraction of kinetic traces every 10 nm, between 420 and 770 nm.
2) Individual fitting of a few sampled traces to obtain (i) an accurate fitting equation and (ii) a good initial guess for the global fit (faster convergence).
3) Performing of the global fit:
   a. Select the best fitting function (from step 2).
   b. Link the time constant(s) for all the kinetic traces included.
   c. Fill up fitting coefficients for each trace according to the guesses obtained in step 2.
   d. Perform fit with a reasonable number of iterations (typically 40):
      i. If convergence is reached: assess fit quality with the calculated residuals.
      ii. If convergence is not reached: readjust initial guesses and repeat.
4) Extraction of the amplitude coefficient(s) for each trace into new sets of data (one set per coefficient, *ie.* per exponential). In our case: $A_1$, $A_2$ and $A_3$ for $\lambda_{ex}$ = 390 nm.
5) Plotting of the constituted amplitude datasets versus wavelengths.

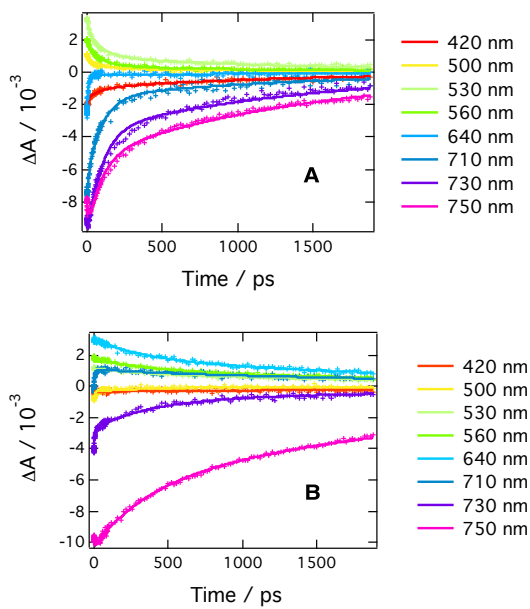

**Figure S1.** Sample of a triexponential global fit results for transient absorption data recorded at $\lambda_{ex}$ = 390 nm. A) Mixed MA$_y$FA$_{1-y}$Br$_x$I$_{3-x}$ perovskite thin film. B) Standard MAPbI$_3$ perovskite thin film.

## 3. Fitting of Equations

Dynamic traces have been fitted with a tri-exponential function:

$$\Delta A(\lambda, t) = y_0 + A_1(\lambda) \cdot \exp\left(\frac{-t}{\tau_1}\right) + A_2(\lambda) \cdot \exp\left(\frac{-t}{\tau_2}\right) + A_3(\lambda) \cdot \exp\left(\frac{-t}{\tau_3}\right) \quad (S2)$$

As the datasets reported in Figs S2(A) and S2(B) span a time window between 2 ps and 1880 ps, no convolution with the instrument response function (IRF) was needed (Gaussian IRF with FWHM= 250 fs $\lambda_{ex}$ = 390 nm).

## 4. Current Density-Voltage (*J-V*) Curves

The data presented in Table 1 of the main text are based on the following *J-V* (Fig. S3) and maximum power point tracking (Fig. S4) measurements, on 5×5 mm square solar cells. Note that the *J-V* data have been obtained for both backward and forward scan directions with a scan rate of 10 mV/s, in air and at ambient temperature.

A xenon lamp equipped with AM1.5 G filter from Newport was used as the light source, together with a silicon reference diode equipped with a KG3 filter. A 4×4 mm (0.16 cm$^2$) black mask was employed throughtout the measurements.

It is important to note that PCE values presented in Fig. S3 are relatively low compared to the figures reported for the very best solar cells made of the same materials and fabricated in similar ways. These comparatively low efficiencies are actually due to the use of thin perovskite layers, whose absorbance is not sufficient for harvesting totally the incident light. The thickness of the active layer was limited to match that of the samples used for spectroscopic measurements.

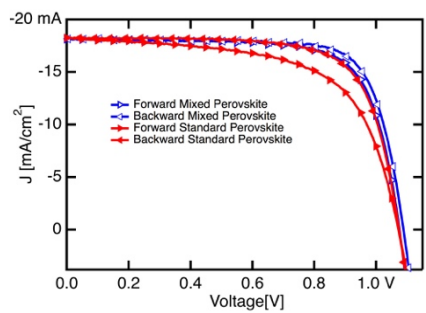

**Figure S2.** Backward and forward scanned *J-V* curve recorded for test solar cell elements (0.16 cm$^2$) made of standard and mixed perovskite films.

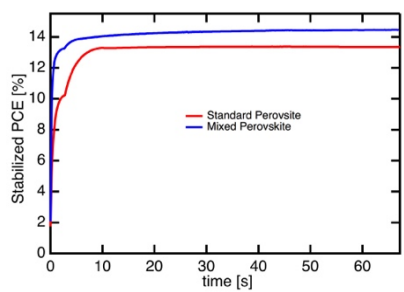

**Figure S3.** Maximum power point tracking measured for test solar cell elements made of standard and mixed perovskite films.

## References

1 G. Lanzani, "The Photophysics Behind Photovoltaics and Photonics". Wiley-VCH; Weinheim, 2012.
2 I. H. M. Van Stokkum, D. S. Larsen, R. van Grondelle, *Biochim. Biophys. Acta Bioenerg.* 2004, **1657**, 82-104.
3 R. Berera, R. van Grondelle, J. T. M. Kennis, *Photosynth. Res.* 2009, **101**, 105-118.
4 M. E. F. Bouduban, A. Burgos-Caminal, R. Ossola, J. Teuscher, J.-E. Moser, *Chem. Sci.* 2017, **8**, 4371-4380.